\documentclass[prl,twocolumn,notitlepage,longbibliography,amsmath,amssymb,floats,superscriptaddress,nofootinbib,10pt]{revtex4-2}
\usepackage{graphicx,color}
\usepackage{amsmath}
\usepackage{amsfonts, amsbsy}
\usepackage{amssymb}
\usepackage{eqnarray}
\usepackage{bm}
\usepackage{blkarray}
\usepackage{mathtools}
\usepackage{psfrag}
\usepackage{caption}
\usepackage{subcaption}
\usepackage{multirow}
\usepackage{textcomp}
\usepackage{units}
\usepackage{lipsum}
\usepackage[title]{appendix}
\usepackage{soul}
\usepackage{titlesec}
\usepackage{times}
\usepackage{enumitem}
\usepackage{soul}
\usepackage[sort&compress]{natbib}
\usepackage[breaklinks,colorlinks = true,linkcolor = blue,urlcolor  = blue,citecolor = blue,anchorcolor = blue]{hyperref}
\usepackage{float}
\usepackage{orcidlink}

\usepackage{xcolor}

\begin{document}
\title{Non-equilibrium charge-vortex duality}

\author{Lazaros Tsaloukidis \orcidlink{0000-0003-1292-1037}}
\email{ltsalouk@pks.mpg.de}
\affiliation{Max Planck Institute for the Physics of Complex Systems, N\"othnitzer Str. 38, 01187, Dresden, Germany}
\affiliation{W\"urzburg-Dresden Cluster of Excellence ct.qmat, 01187, Dresden, Germany}

\author{Francisco Pe\~na-Ben\'itez \orcidlink{0000-0001-8830-2940}}
\affiliation{Institute of Theoretical Physics, Wroc\l{}aw  University  of  Science  and  Technology,  50-370  Wroc\l{}aw,  Poland}

\author{Piotr Sur\'owka \orcidlink{0000-0003-2204-9422}}
\affiliation{Institute of Theoretical Physics, Wroc\l{}aw University of Science and Technology, 50-370 Wroc\l{}aw, Poland}
\affiliation{Institute of Condensed Matter Physics, Department of Physics, Technical University of Darmstadt, Hochschulstr. 8,
64289, Darmstadt, Germany}

\date{\today}

\begin{abstract}
 Traditionally applied within equilibrium states, the charge-vortex dualities are expanded to address the complex dynamics of superfluids and ideal fluids under non-static conditions. We have constructed explicit mappings of finite temperature fluid dynamics to gauge theories, enabling a dual description where vortices in both superfluids and ideal fluids are interpreted as charges within these theories. We found that vortices in the normal component naturally exhibit mobility restrictions, as manifested by the symmetries and conservation laws. Next, we formulated the Liénard-Wiechert problem for the ideal fluid at finite temperature and extracted the wave dynamics in the system along with the speed of sound corrections for both fluid components. Finally, we computed the correlation functions for the gauge potentials, particularly elucidating explicit cross-correlations between the normal and superfluid components.
\end{abstract}

\maketitle 

Charge-vortex dualities in two-dimensional systems provide a fascinating perspective on the symmetries and dynamics that govern the behavior of fields and particles in a planar geometry \cite{savit_duality_1980,senthil_duality_2019,grosvenor_space-dependent_2022}. This concept fundamentally arises from the theoretical framework of dualities in field theories, which typically express one physical scenario in terms of another, often leading to profound insights about the behavior of systems under varying conditions, such as temperature and coupling constants.

Charge-vortex dualities enable transformations between theories describing charged particles and those describing vortices, offering a unique window into understanding phenomena such as superconductivity and the quantum Hall effect. The principle here hinges on representing the interactions or dynamics of one type of excitation the viewpoint of another, fundamentally different type, thus revealing deeper structures of the physical system.

The real power of charge-vortex dualities emerges in their ability to simplify complex physical problems. For example, in situations where the behavior of vortices in a system is difficult to compute due to strong coupling effects, the dual theory in terms of charges might be weakly coupled and hence computationally tractable. This inversion of complexity not only aids theoretical calculations but also helps in providing qualitative insights into phase transitions, critical behavior, and topological properties of materials. Such dualities are not just theoretical constructs but have practical implications in designing experiments and interpreting observational data in condensed matter physics, offering a robust toolset for exploring new states of matter in two-dimensional materials.

In superfluids and superconductors, the charge-vortex duality is particularly crucial \cite{peskin_mandelstam-t_1978,dasgupta_phase_1981,lee_anyon_1991}. It describes a transformation where vortices, topological defects in the phase field of a superfluid or a superconductor, are mapped to charges in a dual $U(1)$ electromagnetic field. For superconductors, which can be seen as charged superfluids of Cooper pairs, this duality is key to understanding electromagnetic interactions under magnetic fields. The concept is also extended in elasticity theories, where it maps elastic defects to charges in tensor gauge theories, thus enhancing our understanding of material responses under stress \cite{kleinert_duality_1982,kleinert_double_1983,beekman_dual_2017-1,beekman_dual_2017,pretko_fracton-elasticity_2018,pretko_crystal--fracton_2019,kumar_symmetry-enforced_2019,radzihovsky_fractons_2020,radzihovsky_quantum_2020,gromov_duality_2020,nguyen_fracton-elasticity_2020,surowka_dual_2021,gaa_fracton-elasticity_2021,hirono_effective_2022,caddeo_emergent_2022,zaanen_crystal_2022,tsaloukidis_fracton-elasticity_2024,nguyen_quantum_2024,Glodkowski:2024qsf,du_quantum_2024}. Traditionally, these dualities have been applied to systems in equilibrium, as the effective actions for spontaneously symmetry-broken condensates are predominantly understood in these conditions.

Recent developments in non-equilibrium physics have facilitated the extension of effective field theories to dynamic, non-static scenarios. These developments include the coset construction for Goldstone bosons within the Schwinger-Keldysh formalism, which systematically develops low-energy effective actions for Nambu-Goldstone modes under non-equilibrium conditions. This method starts by identifying the symmetry breaking patterns and applying the Maurer-Cartan form to derive the corresponding actions. It also incorporates specific adjustments for non-equilibrium scenarios, such as the dynamical Kubo-Martin-Schwinger conditions, ensuring that the effective actions align with the statistical and thermal dynamics of the system \cite{landry_coset_2020,akyuz_schwinger-keldysh_2024}.

Adopting the Schwinger-Keldysh formalism effectively doubles the field content to accommodate both forward and backward time evolution on the closed-time path, capturing the full dynamics of mixed states and maintaining the causality and unitarity essential to the physical system. This dual-field setup is pivotal for describing temporal evolutions and interactions within these complex systems \cite{sieberer_keldysh_2016,liu_lectures_2018}.

In this paper we construct explicit mappings between finite temperature ideal fluids and superfluids, leading to formulations of the effective action for the Euler equations and the Landau-Tisza model as gauge theories. In these theories, vortices are mapped to the charges of gauge theories, broadening our perspective on fluid dynamics and phase transitions in both equilibrium and non-equilibrium frameworks.

\section{Correspondence principle between elasticity and hydrodynamics} As a motivation for the development of dual gauge theory description for fluids we start with the correspondence that emerges between equations of linear elasticity and the Stokes' equations in hydrodynamics \cite{pipkin_lectures_1986,furst_microrheology_2017}.

For incompressible viscous liquids, the equations are:
\begin{align*}
\nabla \cdot \dot{u} &= 0, \\
\rho \ddot{u} &= -\nabla p + \eta \nabla^2 \dot{u},
\end{align*}
illustrating how viscous fluids respond to internal and external pressures and viscous forces, emphasizing their dissipative nature.

Conversely, isotropic, incompressible elastic solids are described by:
\begin{align*}
\nabla \cdot u &= 0, \\
\rho \ddot{u} &= -\nabla p + G \nabla^2 u,
\end{align*}
highlighting the elastic restoring forces characteristic of solids, contrasting with the viscous damping seen in fluids.

Under oscillatory forces the equations become:
\begin{align*}
-\rho \omega^2 \tilde{u} &= -\nabla \tilde{p} + i \omega \eta \nabla^2 \tilde{u}, \\
-\rho \omega^2 \tilde{u} &= -\nabla \tilde{p} + G \nabla^2 \tilde{u},
\end{align*}
with the key distinction lying in the nature of the shear modulus: imaginary for fluids and real for solids.
We see that the form of elastic and linear fluid equations is exactly the same provided that we substitute elastic coefficients with viscosities. Therefore we conclude that upon the above substitution the dual gauge theory equations in elasticity will describe Stokes flows. Although constructing explicit maps beyond the linear regime is more challenging (see \cite{shinn_electrodynamics_2025} for an example of a non-linear map), we intend to build a systematic procedure for the dual description of fluids and finite-temperature superfluids based on effective actions.

\section{Effective theory of superfluids at finite temperature} 
We consider a superfluid characterized by its phase field, $\psi$, which signifies the Goldstone mode resulting from $U(1)$ symmetry breaking. Concurrently, we include analogous fields relevant for the description of the normal component to describe interactions within the two-fluid medium. In a setting that is both isotropic and homogeneous, the description of these fields is presented as follows:
\begin{subequations}\label{eq:Maxwelle}
\begin{equation} 
\psi(x) \propto t + \pi^0(x),
\end{equation}
\begin{equation} 
\phi^I(x) \propto x^I + \pi^I(x),
\end{equation}
\end{subequations}
where $\pi^0(x)$ and $\pi^I(x)$ represent fluctuations around base states of the superfluid phase and the positions of the normal component, respectively. The analysis is conducted in Lagrangian coordinates, which track the motion of fluid elements as they move through space (see \cite{soper_classical_2008,bennett_lagrangian_2006} for a reviews). This approach contrasts with Eulerian descriptions, which focus on specific locations in space through which the fluid flows. The use of Lagrangian coordinates allows us to describe the fluid more naturally in terms of its intrinsic properties, considering how individual elements of the fluid evolve over time.

To analyze the dynamics under small perturbations, one expands the Lagrangian to second order in the perturbations $\pi^0$ and $\vec{\pi}$, resulting in the following quadratic Lagrangian describing relativistic fluids \cite{Nicolis:2011cs,landry_coset_2020}
\begin{equation}
\label{Eq:eft}
\begin{split}
{\cal L}  = & \frac{1}{2} \big[ K_N  \dot{\vec \pi}^2 - G_N (\vec \nabla \cdot \vec \pi)^2 \big] \\
   & + \frac{1}{2} \big[ K_S( \dot \pi^0)^2 - G_S (\vec \nabla \pi^0)^2   \big] + M \, (\vec \nabla \cdot \vec \pi) \dot \pi^0 .
\end{split}
\end{equation}
Here, $K_N$, $G_N$, $K_S$, and $G_S$ denote the kinetic and gradient energies for the normal and superfluid components, while $M$ represents the mixing term between the fluid motions. Below we introduce $D_{ijkl}=G_N\delta_{ij}\delta_{kl}$ in order to make the analogy with elasticity explicitly.
This formulation is particularly useful in formulating the dual representation of fluids and superfluids as we will argue below. The fluid four-velocity is defined as
\begin{equation}
u^\mu = \frac{1}{b} \mathcal{J}^\mu \; ,
\end{equation}
where $\mathcal{J}^\mu = \frac{1}{2} \epsilon^{\mu\alpha\beta} \epsilon_{IJ} \partial_\alpha \phi^I \partial_\beta \phi^J $ and $b = \sqrt{-\mathcal{J}_\mu \mathcal{J}^\mu}=b_0[1+\partial_b\pi^b+\mathcal{O}(\partial^2)] $. Fluid velocity and temperature to this order of expansion read
\begin{subequations}
\begin{align}
u^0 &= 1 + \mathcal{O}(\partial^2), \\
u^i &= -\partial_t\pi^i  + \mathcal{O}(\partial^2), \\
T(b) &= T(b_0) + \left.\frac{\partial T}{\partial b}\right|_{b_0}(b-b_0) + \mathcal{O}(\partial^2).
\end{align}
\end{subequations}

\section{Dual formulation} In exploring the effective action description of superfluids, we shift our focus toward developing a dual formulation. This step is motivated by recognizing that boson-vortex type dualities fundamentally rely on an effective action that portrays a Goldstone field, which may exhibit singular behavior. To handle these cases, we utilize canonical variables that impose a set of conservation-like constraints. These constraints are subsequently addressed through the introduction of gauge fields, a strategy that has proven effective in the study of equilibrium superfluids. In these contexts, Goldstone fields arise from the $U(1)$ symmetry and are similarly employed in elastic theories, where they represent various space-time symmetries.

Despite the parallels between elasticity and fluid dynamics, the inherently non-equilibrium nature of fluids has hindered the application of this dual approach. We identify the primary obstacle as the traditional Eulerian view, which characterizes fluids in terms of velocities. Unfortunately, these fields are not suitable for forming effective actions in fluid dynamics. To overcome this limitation and align fluid dynamics with elasticity under a unified theoretical framework, it becomes essential to adopt Lagrangian fields. This shift is encapsulated in the action defined by Eq. \eqref{Eq:eft}, paving the way for a coherent duality between elasticity and fluid dynamics. We see that setting the superfluid component to zero we arrive at the action that corresponds verbatim to the action of elasticity with shear transport coefficients vanishing. As a result we can now follow the standard procedure of the boson-vortex dualities and formulate both finite temperature ideal fluids and superfluids as gauge theories. This elasticity-fluid dynamics correspondence will be our guiding principle throughout this paper. 

We start by introducing the dual variables
\begin{equation}
\begin{split}
&\Sigma_0=K_S(\partial_t\pi^0), \qquad \Sigma_i=G_S(\partial_i\pi^0),\\ &P_i=K_N(\partial_t\pi_i), \qquad T_{ij}=D_{ijkl}(\partial_k\pi_l),
\end{split}
\end{equation}
where $\Sigma_0$ and $\Sigma_i$ corresponds to the canonical momentum density and a current associated with superfluid velocity. $P_i$ is the canonical momentum density and $T_{ij}$ the current of the normal component. In the analysis of fluctuation fields within the theoretical framework of boson-vortex dualities, it is instructive to decompose these fields into their respective smooth and singular components. Following this separation, the next critical step involves integrating out the smooth part of the fluctuation fields. This leads to the following constraints
\begin{equation}
\begin{split}
&\partial_t\Sigma_0-\partial_i\Sigma_i+\frac{M}{2G_N}\partial_tT_{ii}=0,\\ &\partial_tP_i-\partial_jT_{ij}+\frac{M}{K_S}\partial_i\Sigma_0=0 ,   
\end{split}    
\end{equation}
(for a detailed derivation see Appendices). In analogy with fracton-elasticity dualities these constraints can be resolved by the introduction of gauge fields
\begin{equation}
\begin{split}
&\mathcal{B}=\Sigma_0+\frac{M}{2G_N}T_{ii} \qquad \mathcal{E}_{i}=\epsilon_{ij}\Sigma_j\\
&B_i=\epsilon_{ij}P_j \qquad E_{ij}=\epsilon_{ik}\epsilon_{jl}T_{kl}-\frac{M}{K_S}\delta_{ij}\Sigma_0,
\end{split}
\end{equation}
which in turn can be used to formulate the dual gauge theory incarnation of the ideal (relativistic) Landau-Tisza superfluids and Euler fluids:
\begin{equation}
\label{eq:gaugedual}
\begin{split}
S_{\text{dual}}=&\frac{1}{2}\int d^2xdt\left[G_S^{-1}\mathcal{E}_i^2-\frac{1}{C_1}\mathcal{B}^2+\tilde{D}^{-1}_{ijkl}E_{ij}E_{kl}\right.\\
&\left.-K_N^{-1}B_i^2+\frac{1}{C_3}E_{ii}\mathcal{B}\right].
\end{split}
\end{equation}
where in the above, we set the coefficients
\begin{equation}
\frac{1}{C_1}=\frac{G_N}{G_N K_S + M^2}, \hspace{1cm} 
\frac{1}{C_3}=\frac{M}{G_N K_S + M^2}
\end{equation}
and the inverted modified longitudinal tensor $\tilde{D}^{-1}_{ijkl}=\dfrac{K_S}{G_N K_S + M^2}\dfrac{\delta_{ij} \delta_{kl}}{4}.$

\section{Defects/vortices} The effective field theory we have developed so far overlooks a crucial component of fluid dynamics—vortices. Vortices represent singular configurations in fluid flow, characterized by large-scale rotational movements. These formations inherently break the initial symmetries of the theory by establishing a preferred rest frame for the vortices, which introduces spontaneous symmetry breaking.

In the context of dual theories, these singular configurations are particularly significant as they act like charges for the gauge fields. As we shall see, this dual formulation explicitly highlights the symmetries inherent to vortices, shedding light on their unique properties and behaviors. 

Vortex motion on a plane is a classic subject \cite{saffman_vortex_2012,aref_point_2007}. Motivated by detailed studies of point-vortex dynamics on a plane, it is evident that vortices typically display constrained dynamics. This behavior arises from the conservation of both the dipole moment and the trace of the quadrupole moment \cite{chapman_ideal_1978,doshi_vortices_2021}. The unconventional symmetry properties of vortices have motivated effective field theory developments in the studies of their macroscopic, many-body properties \cite{onsager_statistical_1949,bradley_energy_2012,wiegmann_anomalous_2014,lucas_sound-induced_2014,yu_emergent_2017,shinn_electrodynamics_2025}. 

This observation closely mirrors phenomena observed in elasticity, where topological defects exhibit restricted or "fractonic" motion. Such parallels enhance our understanding of the underlying principles governing both fluid dynamics and elasticity, suggesting a deeper intrinsic linkage between the dynamics of vortices and the behavior of elastic materials with topological constraints. 

In analogy to elasticity, vortices are described as singular configurations of fields in our theory. Upon performing the duality, vortices become charges of the dual gauge fields
\begin{equation}
S_{\text{defect}}=\int d^2xdt\left(-\rho\phi+A_iJ_i-\tilde{\rho}\psi+A_{ij}\tilde{J}_{ij} \right)    ,
\end{equation}
where the explicit formulae for the densities and currents take the form
\begin{equation}
\begin{split}
&\rho=\epsilon_{ij}\partial_i\partial_j\pi^{0(s)}, \quad J_i=\epsilon_{ij}(\partial_j\partial_t-\partial_t\partial_j)\pi^{0(s)},
\end{split}  
\end{equation}
\begin{equation}
\begin{split}
&\tilde{\rho}=\epsilon_{ik}\epsilon_{jl}\partial_k\partial_l\partial_i\pi_j^{(s)}, \quad \tilde{J}_{ij}=\epsilon_{ik}\epsilon_{jl}(\partial_k\partial_t-\partial_t\partial_k)\pi_l^{(s)},
\end{split}  
\end{equation}
(for a detailed derivation see Appendices). In a superfluid at finite temperature, we observe two distinct species of vortices. One species exists within the superfluid condensate and is described by the $U(1)$ gauge field. The other species corresponds to the normal component of the fluid, which sources the symmetric tensor gauge field. Notably, vortices in the normal component face mobility constraints if the number of vortices is fixed and cannot fluctuate.\footnote{We note that vortices are not simply related to vorticity field $\omega = \epsilon^{\alpha\beta\gamma} u_\alpha \nabla_\beta u_\gamma.
$}
Physically, the above constraint means that while a single vortex on a plane cannot move, a dipole of vorticity is capable of movement. This observation parallels the symmetry structure seen in point vortex dynamics. However, it's important to note that in the previous context, the symmetry exists at the level of solutions. In our scenario, the mobility restrictions are inherent to the theory itself. Additionally, we begin with a relativistic effective action, indicating that the symmetry of the state with vortices is analogous to its non-relativistic counterpart. We observe that the singularities within the superfluid component do not exhibit constrained dynamics according to the theory itself. This can be readily seen by looking at the conservation equations
\begin{equation}
\partial_t\rho+\partial_iJ_i=0, \qquad \partial_t\Tilde{\rho}+\partial_i\partial_j\tilde{J}_{ij}=0.  
\end{equation}
Superfluid vortices obey conventional continuity equations, whereas the continuity for the normal component singularities is of the fractonic type. In turn, the point-vortex dynamics is not the most general approach to model superfluid vortices as the symmetries do not match.  

Moreover, it is crucial to note that the descriptions of the ideal normal component and the superfluid component differ significantly. This distinction is important as it contrasts with the conventional approach where ideal fluids and superfluids are often treated as similar. The difference can be traced to the different symmetry breaking pattern and associated Goldstone degrees of freedom.

In order to see explicitly the macroscopic behavior of moving vortices it is convenient to formulate the Liénard–Wiechert problem for our theory generalizing \cite{radzihovsky_anomalous_2015,tsaloukidis_elastic_2024} (for a detailed derivation see Appendices). This amounts to writing down the equations for the evolution of gauge potentials in the presence of the sources in the Lorenz gauge. In our approach, we utilize the field $\psi_i = \partial_i \psi $, and additionally, we introduce the dipole density $\rho_i$ for the vortices of the normal component. For this dipole density, we have $
\partial_i \tilde{\rho}_i = \tilde{\rho},$
mirroring the scenario observed in solids. We then extract the wave velocities for normal and superfluid components
\begin{subequations}
\begin{equation}
\upsilon_{N}\approx\sqrt{\frac{G_N}{K_N}}\left(1-\frac{M^2}{2G_NK_S}\right) ,
\end{equation}
\begin{equation}
\upsilon_{S}\approx\sqrt{\frac{G_S}{K_S}}\left(1-\frac{M^2}{2G_NK_S}\right),
\end{equation}
\end{subequations}
in the limit of small coupling $M$.

{\it Correlation functions} --- In systems undergoing spontaneous symmetry breaking, the correlation functions play a pivotal role in elucidating the emergent physical properties and dynamics. 

For instance, in quantum field theories describing quantum crystals, they provide insights into how topological defects, such as vortices or disclinations, influence the field configurations over large distances. The long-range correlations established by these defects are essential for the formation of ordered phases from disordered states, as seen in dual gauge theory. These topological defects act as sources for gauge fields, with the correlation functions describing how the field disturbances decay or persist across the system.

Moreover, their behavior near critical points, where the system transitions from one phase to another, can reveal the nature of the phase transition, such as its order and critical exponents. This is particularly relevant in quantum phase transitions, where temperature does not play a role, but quantum fluctuations drive the transition.

We compute the analogues of photon propagators in the gauge theory given by Eq. \eqref{eq:gaugedual} in the limit of small coupling $M$. These depend on the gauge choice, however, we can use them, to compute gauge invariant correlation functions of the electric and magnetic fields. By making use of the isotropy of the system and fixing the fluctuations only along the $x$ axis, in the Coulomb gauge we get
\begin{subequations}\label{eq:corrcross}
\begin{equation} 
<A_y,\psi>=-\frac{iC_2}{C_3k}\frac{G_S\left(1+\frac{\omega^2k^2}{C_3^2}\frac{1}{f_1(\omega,k)f_2(\omega,k)}\right)}{\omega^2-(G_S/C_1)k^2},
\end{equation}
\begin{equation} 
<A_y,A_{yy}>=-\frac{\omega k}{C_3}\frac{\left(1+\frac{\omega^2k^2}{C_3^2}\frac{1}{f_1(\omega,k)f_2(\omega,k)}\right)}{f_1(\omega,k)f_2(\omega,k)}.
\end{equation}
\label{eq:crosscorr}
\end{subequations}
where we used $f_1(\omega,k)=\left(\frac{\omega^2}{C_2}-\frac{k^2}{K_N}\right)$ and $f_2(\omega,k)=\left(\frac{\omega^2}{G_S}-\frac{k^2}{C_1}\right)$ for shortness (for a detailed derivation see Appendices). Finally we note that in a two-fluid mixture there is no static force between vortices in the superfluid component and the normal component. However, there is a dynamical coupling once vortices have moved. Specifically, the mixed correlators $\langle A_y, \psi \rangle$ and $\langle A_y, A_{yy} \rangle$ indicate that fluctuations in the superfluid $U(1)$ gauge field $A_y$ are not dynamically independent of the tensor gauge fields associated with the normal component. This interdependence arises due to the coupling term $M$ in the original Lagrangian, and manifests through the $C_3$-dependent contributions in the correlation functions.

Physically, these mixed correlators imply that a moving vortex in one component (e.g., the superfluid) can induce gauge field responses in the other component (e.g., the normal fluid). Although the two-fluid system exhibits no static interaction between vortices residing in different components, the non-vanishing of these mixed two-point functions highlights a non-trivial dynamical coupling.

\section{Discussion} We have provided an example of particle-vortex duality that applies to superfluids and fluids at finite temperature. It thus serves a first example of a non-equilibrium duality of the particle-vortex type that maps strong to weak coupling regimes. In consequence our framework offers a unique opportunity to study properties of fluids at strong coupling. Perhaps it can shed light on the classic problems of fluid mechanics, when further genaralized to include viscous effects.

At the technical level, our paper presents a gauge theory incarnation of fluid dynamics, a classic field in mathematical physics. Such a gauge decomposition is typically achieved only for limited examples using Clebsch potentials \cite{lamb_hydrodynamics_2005,Lin1963,Seliger_variational_1968,scholle_first_2011,tong_gauge_2023}. However, our framework, based on tensor gauge theories, does not exhibit the usual limitations and complexities associated with Clebsch variables. Therefore, it may be worthwhile to understand and expand classical results within the language of our framework. Apart form the gauge structures of fluids, recent studies on symmetry breaking mechanisms in fluids, revealed a new pattern for spontaneous symmetry breaking \cite{sala_spontaneous_2024,lessa_strong--weak_2025,huang_hydrodynamics_2025}. It would be interesting to elucidate it further in our framework.

Finally, a question has emerged as to whether fluids exist at zero temperature and can be quantized analogously to superfluids \cite{endlich_quantum_2011, gripaios_quantum_2015, wiegmann_quantum_2019, dersy_quantum_2024,Cuomo:2024ekf}. This work offers a new perspective on the problem by allowing the quantization of a gauge theory.

{\it Acknowledgements} --- 
This work was supported in part by
the Deutsche Forschungsgemeinschaft under cluster of excellence
ct.qmat (EXC 2147, Project-ID No. 390858490), the Polish National Science Centre (NCN) Sonata Bis grant 2019/34/E/ST3/00405, and the Unite! University alliance.

\bibliography{bibliography.bib}%

\appendices

\section{APPENDIX A: DUAL ACTION}
We start with the action 
\begin{equation}
\begin{split}
S=&\frac{1}{2}\int d^2xdt\left[K_N(\partial_t\pi_i)^2-D_{ijkl}(\partial_i\pi_j)(\partial_k\pi_l)\right.\\
&\left.+K_S(\partial_t\pi^0)^2-G_S(\partial_i\pi^0)^2+2M(\partial_i\pi_i)(\partial_t\pi^0)\right]   ,  
\end{split} 
\end{equation}
where $D_{ijkl}=G_N\delta_{ij}\delta_{kl}$, and we transform it into the dual representation through the functional
\begin{equation}
\mathcal{Z}=\int D[\Phi]e^{-S_{eff}}   
\end{equation}
with $D[\Phi]=D[\partial_t\pi_i,\partial_i\pi_j,\partial_t\pi^0,\partial_i\pi^0]$ being the measure of the fields representing energy densities of the normal and superfluid sectors respectively. By completing the square, we end up with a renormalization factor and the relative Hubbard-Stratonovich fields of the dual action
\begin{equation}
\begin{split}
&S_{\text{dual}}=\frac{1}{2}\int d^2xdt\left[G_S^{-1}\Sigma_i^2-K_S^{-1}\Sigma_0^2+D_{ijkl}^{-1}T_{ij}T_{kl}\right.\\
&-K_N^{-1}P_i^2-\frac{2M}{K_SG_N}T_{ii}\Sigma_0+2P_i(\partial_t\pi_i)+2\Sigma_0(\partial_t\pi^0)\\
&\left.-2\Sigma_i(\partial_i\pi^0)-2T_{ij}(\partial_i\pi_j)+\frac{M}{G_N}T_{ii}(\partial_t\pi^0)+\frac{2M}{K_S}\Sigma_0(\partial_i\pi_i)\right],
\end{split}
\end{equation}
where the constitutive relations between the original variables and the dual ones are:
\begin{equation}
\begin{split}
&\Sigma_0=K_S(\partial_t\pi^0), \qquad \Sigma_i=G_S(\partial_i\pi^0),\\ &P_i=K_N(\partial_t\pi_i), \qquad T_{ij}=D_{ijkl}(\partial_k\pi_l),
\end{split}
\end{equation}
with $P_i$ being the momentum density, $T_{ij}$ the symmetric longitudinal stress tensor of the normal component and $D_{ijkl}^{-1}=\dfrac{1}{G_N}\dfrac{\delta_{ij}\delta_{kl}}{4}$. Separation of the fluctuation fields into a smooth and singular part (denoted by (s)), and integrating out the smooth part, leads to the following linearly coupled equations of motion
\begin{equation}
\begin{split}
&\partial_t\Sigma_0-\partial_i\Sigma_i+\frac{M}{2G_N}\partial_tT_{ii}=0,\\ &\partial_tP_i-\partial_jT_{ij}+\frac{M}{K_S}\partial_i\Sigma_0=0  .  
\end{split}    
\end{equation}
In our case the fluid is taken to be isotropic, which means the stress tensor is related with the fluid pressure through $T_{ij}\sim(\delta_{ij}/2)P$.
\section{APPENDIX B: GAUGE THEORY AND EM FIELDS}
To establish an electromagnetic framework, we refine our dual theory by transforming the previously introduced variables for each component of the fluid. By rearranging the final terms in the equations of motion, we effectively derive the Faraday laws pertinent to the dual gauge theory
\begin{equation}
\partial_t\left(\Sigma_0+\frac{M}{2G_N}T_{ii}\right)-\partial_i\Sigma_i=0 \longrightarrow \partial_t\mathcal{B}+\epsilon_{ij}\partial_i\mathcal{E}_j=0    
\end{equation}
and
\begin{equation}
\partial_tP_i-\partial_j\left(T_{ij}-\frac{M}{K_S}\delta_{ij}\Sigma_0\right)=0\longrightarrow \partial_tB_i+\epsilon_{jk}\partial_jE_{ki}=0,   
\end{equation}
where the electric and magnetic fields are given from
\begin{equation}
\begin{split}
&\mathcal{B}=\Sigma_0+\frac{M}{2G_N}T_{ii}, \qquad \mathcal{E}_{i}=\epsilon_{ij}\Sigma_j,\\
&B_i=\epsilon_{ij}P_j, \qquad E_{ij}=\epsilon_{ik}\epsilon_{jl}T_{kl}-\frac{M}{K_S}\delta_{ij}\Sigma_0.
\end{split}
\end{equation}
The vector electric field $\mathcal{E}_i$ and the scalar magnetic $\mathcal{B}$ of the superfluid represent a regular U(1) sector in 2+1 dimensions, while the higher rank electric and magnetic fields of the normal component are similar to the ones introduced for the description of a solid in said dimensions, albeit with one less degree of freedom, since there is only longitudinal motion for that component of the fluid.\\
The solutions of the Faradays laws above can be found by introducing the gauge fields
\begin{equation}
\mathcal{B}=\epsilon_{ij}\partial_iA_j, \qquad \mathcal{E}_i=-\partial_tA_i-\partial_i\phi,
\end{equation}
\begin{equation}
B_i=\epsilon_{jk}\partial_jA_{ki}, \qquad E_{ij}=-\partial_tA_{ij}-\partial_i\partial_j\psi   .
\end{equation}
Using all the definitions until now, we can arrive at the final action for the EM fields
\begin{equation}
\begin{split}
S_{EM}=&\frac{1}{2}\int d^2xdt\left[G_S^{-1}\mathcal{E}_i^2-\frac{1}{C_1}\mathcal{B}^2+\tilde{D}^{-1}_{ijkl}E_{ij}E_{kl}\right.\\
&\left.-K_N^{-1}B_i^2+\frac{1}{C_3}E_{ii}\mathcal{B}\right],
\end{split}
\end{equation}
where in the above the coupling constants are 
\begin{subequations}
\begin{align}
\frac{1}{C_1} &= \frac{G_N}{G_N K_S + M^2}, \\
\frac{1}{C_3} &= \frac{M}{G_N K_S + M^2}, \\
\tilde{D}^{-1}_{ijkl} &=\frac{K_S}{G_N K_S + M^2}\frac{\delta_{ij} \delta_{kl}}{4}.
\end{align}
\end{subequations}
The two sectors decoupled from one another in the limit where $M\rightarrow 0$, and the constant reduced to the inverted values of the original ones. The new values for them will effectively change the two speed of sounds for the normal and superfluid components as will become apparent later on in the momentum space analysis. 
\section{APPENDIX C: GAUGE TRANSFORMATIONS AND CONTINUITY EQUATIONS}
The gauge transformations keeping the gauge fields invariant are
\begin{equation}
\delta A_i=\partial_if, \quad \delta\phi=-\partial_tf, \quad \delta A_{ij}=\partial_i\partial_jg, \quad \partial\psi=-\partial_tg  ,  
\end{equation}
where $f(x,t)$ and $g(x,t)$ are some arbitrary scalar functions. Introducing matter fields in the action we can extract the continuity equations by performing the gauge transformation of the fields sourced by the charges:
\begin{equation*}
\delta S_{\text{source}}=\int \left[-\rho(\delta \phi)+J_i(\delta A_i)+\tilde{\rho}(\delta \psi)+\tilde{J}_{ij}(\delta A_{ij})\right]=0,
\end{equation*}
which implies
\begin{equation*}
\delta S_{\text{source}}=\int\left[-\rho(\partial_t f)+J_i(\partial_i f)+\tilde{\rho}(\partial_t g)+\tilde{J}_{ij}(\partial_i\partial_j g)\right]=0.
\end{equation*}
Obtained through integration by parts, we derive:
\begin{equation}
\partial_t\rho+\partial_iJ_i=0 \hspace{2cm} \partial_t\Tilde{\rho}+\partial_i\partial_j\tilde{J}_{ij}=0   
\end{equation}
Focusing now on the singular part of the action, we will be deriving the formulas for the defects and currents of the theory. We have:
\begin{equation*}
\begin{split}
S_{\text{source}}&=\int d^2xdt \left[P_i(\partial_t\pi^{(s)}_i)+\Sigma_0(\partial_t\pi^{0{(s)}})-\Sigma_i(\partial_i\pi^{0{(s)}})\right.\\
&\left.-T_{ij}(\partial_i\pi^{(s)}_j)+\frac{M}{G_N}T_{ii}(\partial_t\pi^{0{(s)}})+\frac{M}{K_S}\Sigma_0(\partial_i\pi^{(s)}_i)\right]
\end{split}    
\end{equation*}
If we plug in the dual variables as functions of the gauge fields and perform a series of integrations by parts we arrive at the form (standard EM action convention)
\begin{equation*}
S_{\text{source}}=\int d^2xdt\left(-\rho\phi+A_iJ_i-\tilde{\rho}\psi+A_{ij}\tilde{J}_{ij} \right)    
\end{equation*}
with
\begin{equation}
\begin{split}
&\rho=\epsilon_{ij}\partial_i\partial_j\pi^{0(s)} \qquad J_i=\epsilon_{ij}(\partial_j\partial_t-\partial_t\partial_j)\pi^{0(s)}\\
&\tilde{\rho}=\epsilon_{ik}\epsilon_{jl}\partial_k\partial_l\partial_i\pi_j^{(s)} \qquad \tilde{J}_{ij}=\epsilon_{ik}\epsilon_{jl}(\partial_k\partial_t-\partial_t\partial_k)\pi_l^{(s)}\\     
\end{split}  
\end{equation}
The superfluid vortex density here, is in principle the vorticity defined with respect to the fluctuations of the phase field of the $U(1)$ sector. 
\section{APPENDIX D: MOMENTUM SPACE ANALYSIS}
We will now transition to working within Fourier space to derive the two-point correlation functions for the gauge field. The complete action is expressed as follows:
\begin{widetext}
\begin{equation}
\begin{split}
S_{\text{EM}}&=\frac{1}{2}\int d^2xdt\left[\frac{1}{G_S}(\partial_tA_i+\partial_i\phi)^2-\frac{1}{C_1}\epsilon_{ij}\epsilon_{kl}\partial_iA_j\partial_kA_l+\tilde{D}^{-1}_{ijkl}(\partial_tA_{ij}+\partial_i\partial_j\psi)(\partial_tA_{kl}+\partial_k\partial_l\psi)-\right.\\
&\Big.\hspace{7cm}-\frac{1}{K_N}\epsilon_{jk}\epsilon_{mn}\partial_jA_{ki}\partial_mA_{ni}+\frac{1}{C_3}(\partial_tA_{ii}+\partial_i^2\psi)\epsilon_{jk}\partial_jA_k\Big]
\end{split}  
\end{equation}
The gauge fields in Fourier space are:
\begin{equation}
\begin{split}
&\phi=\frac{1}{(2\pi)^3}\int d^2kd\omega e^{i(\Vec{k}\Vec{r}-\omega t)}\Tilde{\phi}(\Vec{k},\omega), \qquad  \psi=\frac{1}{(2\pi)^3}\int d^2kd\omega e^{i(\Vec{k}\Vec{r}-\omega t)}\Tilde{\psi}(\Vec{k},\omega),\\
A_i&=\frac{1}{(2\pi)^3}\int d^2kd\omega e^{i(\Vec{k}\Vec{r}-\omega t)}\Tilde{A}_i(\Vec{k},\omega), \qquad A_{ij}=\frac{1}{(2\pi)^3}\int d^2kd\omega e^{i(\Vec{k}\Vec{r}-\omega t)}\Tilde{A}_{ij}(\Vec{k},\omega).
\end{split}
\end{equation}
Utilizing the fact that $\tilde{\phi}(-\Vec{k},-\omega)=\Tilde{\phi}^*(\Vec{k},\omega)$ and the $\delta$-function identity, we derive the matrix representation of the Lagrangian
\begin{equation}
\mathcal{L}=\frac{1}{2(2\pi)^3}
\begin{pmatrix}
\tilde{\phi}\\
\tilde{A}_i\\
\tilde{\psi}\\
\tilde{A}_{ab}\\    
\end{pmatrix}^{\dagger}
\begin{pmatrix}
\frac{k^2}{G_S} & -\frac{\omega k_j}{G_S}  & 0 & 0\\
-\frac{\omega k_i}{G_S} & \left(\frac{\omega^2}{G_S}-\frac{k^2}{C_1}\right)\delta_{ij}+\frac{k_ik_j}{C_1} & \frac{i}{C_3}\epsilon_{ij}k^2k_j & -\frac{1}{C_3}\epsilon_{ij}\omega k_j\delta_{cd}\\
0 & -\frac{i}{C_3}\epsilon_{ij}k^2k_i & \frac{k^4}{C_2} & \frac{i\omega k_ck_d}{C_2}\\
0 & \frac{1}{C_3}\epsilon_{ij}\omega k_i\delta_{ab} & -\frac{i\omega k_ak_b}{C_2} & 
\left[\frac{\omega^2}{C_2}\delta_{ab}\delta_{cd}-\left(\frac{k^2\delta_{ac}-k_ak_c}{K_N}\right)\delta_{bd}\right]\\
\end{pmatrix}
\begin{pmatrix}
\tilde{\phi}\\
\tilde{A}_j\\
\tilde{\psi}\\
\tilde{A}_{cd}\\    
\end{pmatrix},
\end{equation}
\end{widetext}
which we can schematically represent as $\mathcal{L}=\frac{1}{2(2\pi)^3}\Phi^T \mathbf{A}\Phi$. In the above we also set $\frac{1}{C_2}=\frac{K_S}{4(G_NK_S+M^2)}$.

Due to the rotational symmetry of the system, we simplify the analysis by considering wave propagation solely in the $x$-direction. We implement the Coulomb gauge fixing, permissible through gauge invariance, by setting $\partial_i A_i = 0 \Rightarrow kA_x = 0$ and $\partial_i A_{ij} = 0$. This leads to the reduction of the system to a $4 \times 4$ matrix that includes the fields $\phi$, $A_y$, $\psi$, and $A_{yy}$, with its determinant significantly simplifying the characterization of wave propagation in the system:
\begin{equation*}
\det \mathbf{A}=\frac{k^6}{C_2G_S}\left[\left(\frac{\omega^2}{C_2}-\frac{k^2}{K_N}\right)\left(\frac{\omega^2}{G_S}-\frac{k^2}{C_1}\right)-\frac{\omega^2k^2}{C_3^2}\right]
\end{equation*}
The standing modes and the propagating for the fields are already visible from the above expression and we list below all the relevant 2-point function for the gauge fields
\begin{equation}
\begin{split}
&<\phi,\phi> = \frac{G_S}{k^2},\\
&<A_y,A_y> = \frac{G_S}{\omega^2-\frac{G_S}{C_1}k^2}\left(1+\frac{\omega^2k^2}{C_3^2}\frac{1}{f_1f_2}\right),\\ 
&<\psi,\psi> = \frac{C_2}{k^4},\\
&<A_{yy},A_{yy}>=\frac{C_2}{\omega^2-\frac{C_2}{K_N}k^2}\left(1+\frac{C_2\omega^2}{C_3^2k^2}\frac{1}{f_1f_2}\right),\\
&<\psi,A_{yy}>=-\frac{i\omega C_2}{C_3^2}\frac{1}{\left[\left(\frac{\omega^2}{C_2}-\frac{k^2}{K_N}\right)\left(\frac{\omega^2}{G_S}-\frac{k^2}{C_1}\right)-\frac{\omega^2k^2}{C_3^2}\right]}.
\end{split}
\end{equation}
and the mixed sector components
\begin{equation}
\begin{split}
&<A_y,\psi>=-\frac{iC_2}{C_3}\frac{1}{k}\frac{G_S}{\left[\omega^2-(G_S/C_1)k^2\right]}\left(1+\frac{\omega^2k^2}{C_3^2}\frac{1}{f_1f_2}\right),\\
&<A_y,A_{yy}>=-\frac{\omega k}{C_3}\frac{1}{f_1f_2}\left(1+\frac{\omega^2k^2}{C_3^2}\frac{1}{f_1f_2}\right).
\end{split}    
\end{equation}
All the propagating 2-point functions above were simplified by assuming that the coupling constant $M$ is small, so that the $1/(1-x)\approx 1+x$ expansion is in effect, with off diagonal components in the matrix now providing the corrections in formulas. This is clearly visible on the $1/C_3$ dependence too, as setting equal to 0, diagonalizes the matrix and reverts the 2-point to their original decoupled expressions. In all the above we set $f_1=\left(\frac{\omega^2}{C_2}-\frac{k^2}{K_N}\right)$ and $f_2=\left(\frac{\omega^2}{G_S}-\frac{k^2}{C_1}\right)$. We note that although we show an expansion of the correlation functions it is possible to obtain exact expressions. Since they are rather lengthy we choos not to display them.

\section{APPENDIX E: VORTEX DIPOLE - WAVE SOLUTIONS}
We will now focus on deriving the wave equations for the respective gauge fields, with all necessary gauge fixing being performed in the Lorenz gauge for both sectors. It is crucial first to introduce the dipole charge density $\tilde{\rho}_i$ of the vortices in the normal component of the fluid, analogous to the dislocation density in solids. Using similar reasoning, we define it through the relationship $\partial_i\tilde{\rho}_i = \tilde{\rho}$. This will then couple to a higher-rank gauge field $\psi_i = \partial_i\psi$, which will be instrumental in deriving the wave equations. Here, we present the relevant equations of motion from the dual perspective (effectively the Bianchi identities in the original action)
\begin{equation}
\begin{split}
&\partial_i\mathcal{E}_i=G_S\rho,\\
&\epsilon_{ij}\partial_j\mathcal{B}=C_1\left(J_i+\frac{1}{G_S}\partial_t\mathcal{E}_i\right)-\frac{C_1}{2C_3}\epsilon_{ij}\partial_jE_{kk},\\
&\partial_i^2E_{kk}=-C_2\tilde{\rho}-\frac{C_2}{2C_3}\partial_j^2\mathcal{B},\\
&\epsilon_{ik}\partial_kB_j=K_N\left(\tilde{J}_{ij}+\frac{\delta_{ij}}{C_2}E_{kk}\right)+\frac{K_N}{2C_3}\delta_{ij}\partial_t\mathcal{B}.\\
\end{split}    
\end{equation}
The equations are obviously coupled with contributions coming for the coupling $M$. Careful choosing of gauges will lead to the required result. The relevant equations for the gauge fields are
\begin{widetext}
\begin{equation}
\begin{split}
&\partial_t(\partial_iA_i)+\partial_i^2\phi=-G_s\rho\\
&\partial_i(\partial_jA_j)-\partial_k^2A_i+\frac{C_1}{G_s}(\partial_t^2A_i+\partial_t\partial_i\phi)=C_1J_i+\frac{C_1}{2C_3}\epsilon_{ij}\partial_j\left[\partial_tA_{kk}+\partial_k\psi_k\right]\\
&\partial_t(\partial_jA_{kk})+\partial^2\psi_j=C_2\Tilde{\rho}+\frac{C_2}{C_3}\epsilon_{kl}\partial_j\partial_kA_l\\
&\partial_k\partial_i(A_{ki})-\partial_k^2A_{ii}+\frac{2K_N}{C_2}(\partial_t^2A_{kk}+\partial_t(\partial_k\psi_k))=K_N\Tilde{J}_{ii}+\frac{K_N}{C_3}\epsilon_{kl}\partial_k(\partial_tA_l)\\
\end{split}    
\end{equation}
where for the last one we took the trace as only the longitudinal (bulk) mode is present of the fluid. The two Lorenz gauge fixings we impose are
\begin{equation}
\partial_t\psi_k+\frac{C_2}{4K_N}\partial_kA_{ii}=0 , \hspace{2cm} \partial_t\phi+\frac{G_S}{C_1}\left(\frac{2C_3^2-C_1C_2}{2C_3^2}\right)\partial_jA_j=0  .  
\end{equation}
which along with the fact that our system is isotropic, meaning that the stress tensor $T_{ij}$ (i.e. $E_{ij}$ also) has components lying only in the diagonal, transforms the above system of equations into
\begin{equation}
\begin{split}
&\nabla^2\phi-\frac{2C_1C_3^2}{G_S(2C_3^2-C_1C_2)}\partial_t^2\phi=-G_S\rho,\\
&\nabla^2A_i-\frac{2C_1C_3^2}{G_S(2C_3^2-C_1C_2)}\partial_t^2A_i=-C_1\left(J_i+\frac{C_2}{2C_3}\epsilon_{ij}\tilde{\rho}_j\right),\\
&\nabla^2\psi_j-\frac{4K_N}{C_2}\partial_t^2\psi_j=C_2\tilde{\rho}_j+\frac{C_2}{C_3}\epsilon_{kl}\partial_j\partial_kA_l,\\
&\nabla^2A_{ii}-\frac{4K_N}{C_2}\partial_t^2A_{ii}=-2K_N\tilde{J}_{ii}-\frac{2K_N}{C_3}\epsilon_{kl}\partial_k(\partial_tA_l).\\
\end{split}
\end{equation}
\end{widetext}
The first two of the above equations can be easily solved if one chooses appropriate functions for the superfluid charge density and currents. Additionally, we observe that the dipole density of the normal component couples to the superfluid current with a constant proportional to $M$. The analysis of the second sector is somewhat more complex, as it necessitates solving the first part and then utilizing the solution for the gauge field $A_i$ on the right-hand side to derive a modified Green's function for the normal component. We note that the velocities from both the normal and superfluid components receive corrections proportional to the coupling constant. In the limit of small $M$, these corrections are found to be
\begin{equation}
\upsilon_{N}=\sqrt{\frac{G_N}{K_N}\left(1-\frac{M^2}{G_NK_S}\right)}\approx\sqrt{\frac{G_N}{K_N}}\left(1-\frac{M^2}{2G_NK_S}\right) 
\end{equation}
and
\begin{equation}
\upsilon_{S}\approx\sqrt{\frac{G_S}{K_S}}\left(1-\frac{M^2}{2G_NK_S}\right),
\end{equation}
where again we used the same small $M$ expansion as for the 2-point functions above. So we see that in this case, the correctional factor is the same for both of the fluid components.

Regarding the actual mathematical form of the relativistic solutions, we now consider the case of 2+1 dimensional solutions. Traditionally, the use of $\delta$-function for defects leads to several issues concerning retarded potentials. Given that we are working within the 2+1 dimensional plane, the exact form of the solution is distinct from that observed in the 3+1 dimensional case.

The Green's function related to the aforementioned differential equations is expressed as the Heaviside step-function, incorporating a relativistic factor. This inclusion gives rise to the "afterglow" phenomenon, a direct consequence of Huygens' principle not being applicable in 2+1 dimensions and, more generally, in spacetimes with an odd number of dimensions.

This can be observed from the Green's function that describes the equations of motion. In 3+1 dimensions, it is proportional to Dirac's delta function, which results in an instant impulse followed by a rapid cessation of its effect. In contrast, here the Heaviside $\Theta$ function is present, indicating that while there is no contribution at times $t < t' + |\mathbf{r} - \mathbf{r}'| / \upsilon_s$, the pulse emitted precisely at $t = t'$ has a prolonged impact. The denominator acts as an attenuation factor, diminishing gradually at large values of $t$.

The full wave describing this phenomenon can be modeled as a superposition of wave modes with velocity values ranging from 0 to $\upsilon_s$ (velocity of the outermost wave, the so-called Huygens surface). This is known as the "tail" of the Green's function, leading to non-sharp wave propagation, in contrast to the case in (3+1) dimensions.

\end{document}